\documentclass[prd,twocolumn,floatfix,amsmath,nofootinbib,amssymb,floatfix]{revtex4}
\usepackage{graphicx,color,dcolumn,booktabs,bm}
\usepackage{longtable,lscape}
\usepackage{pdfpages}
\usepackage{txfonts}
\usepackage{overpic}
\usepackage{amssymb}
\usepackage{makecell}
\usepackage{indentfirst}
\usepackage{feynmf}   
\usepackage{slashed}  
\usepackage{cases}
\usepackage{color}
\usepackage{multirow}
\usepackage{threeparttable}
\usepackage{epstopdf}
\usepackage{enumerate}
\usepackage{subfigure}
\usepackage{diagbox}
\usepackage{graphicx,color,dcolumn,booktabs,bm}
\usepackage{mathrsfs}
\usepackage{cancel}
\usepackage{float}
\usepackage[colorlinks,
            citecolor=blue,
            anchorcolor=red,
            menucolor=red,
            linkcolor=red,
            filecolor=red,
            runcolor=red,
            urlcolor=blue,
            frenchlinks=red]{hyperref}

\begin{document}
\title{The strong vertices of bottom mesons $B$, $B^{*}$ and bottomonia $\Upsilon$, $\eta_{b}$}
\author{Jie Lu$^{1}$}
\author{Guo-Liang Yu$^{1}$}
\email{yuguoliang2011@163.com}
\author{Zhi-Gang Wang$^{1}$}
\email{zgwang@aliyun.com}
\author{Bin Wu$^{1}$}

\affiliation{Department of Mathematics and Physics, North China
Electric power university, Baoding 071003, People's Republic of
China}
\date{\today}

\begin{abstract}
In this article, the strong coupling constants of vertices $BB\Upsilon$, $BB^{*}\Upsilon$, $B^{*}B^{*}\Upsilon$, $BB^{*}\eta_{b}$ and $B^{*}B^{*}\eta_{b}$ are analyzed in the framework of QCD sum rules. In this work, all possible off-shell cases and the contributions of vacuum condensate terms including $\langle\overline{q}q\rangle$, $\langle\overline{q}g_{s}\sigma Gq\rangle$, $\langle g_{s}^{2}G^{2}\rangle$, $\langle f^{3}G^{3}\rangle$ and $\langle\overline{q}q\rangle\langle g_{s}^{2}G^{2}\rangle$ are considered. The momentum dependent strong coupling constants are first calculated and then are fitted into analytical functions $g(Q^{2})$ which are used to extrapolate into time-like regions to obtain the final values of strong coupling constants. The final results are $g_{BB\Upsilon}=40.67^{+7.55}_{-4.20}$, $g_{BB^{*}\Upsilon}=11.58^{+2.19}_{-1.09}$ GeV$^{-1}$, $g_{B^{*}B^{*}\Upsilon}=57.02^{+5.32}_{-5.31}$, $g_{BB^{*}\eta_{b}}=23.39^{+4.74}_{-2.30}$ and $g_{B^{*}B^{*}\eta_{b}}=12.49^{+2.12}_{-1.35}$ GeV$^{-1}$. These strong coupling constants are important input parameters which reflect the dynamic properties of the interactions among the mesons and quarkonia.
\end{abstract}


\maketitle

\section{Introduction}\label{sec1}

The suppression of $J/\psi$ production in relativistic heavy ion collisions is an important signature to identify the quark-gluon plasma\cite{Matsui:1986dk}. Because of the color screening, the dissociation of $J/\psi$ in the quark-gluon plasma would lead to a reduction of its production. The bottomonia are also sensitive to the color screening, therefore the $\Upsilon$ suppression in heavy ion collisions can also be considered as a signature to identify the quark-gluon plasma\cite{Vogt:1999cu,Rapp:2008tf}. There are already some successful attempts in analyzing the heavy quarkonium absorptions by the effective Lagrangians in meson exchange models\cite{Matinyan:1998cb,Haglin:1999xs,Lin:1999ad,Sibirtsev:2000aw,Lin:2000ke}. And the absorption cross sections can be calculated basing upon the interactions among the mesons and quarkonia, where the strong coupling constants are taken as an important input parameter. On the other hand, accurate determination of the strong coupling constants plays an important role in understanding the effects of heavy quarkonium absorptions in hadronic matter\cite{Casalbuoni:1996pg}. Besides, the coupling constants among the heavy quarkonia and heavy mesons are valuable for us to understand the final-state interactions in the heavy quarkonium decays\cite{Casalbuoni:1996pg,Meng:2008bq}.

The QCD sum rules and the light-cone QCD sum rules are powerful nonperturbative approaches in analyzing the strong coupling constants among the hadrons. In recent years, the strong vertices $DD^{*}\pi$, $D^{*}D_{s}K$, $DD_{s}^{*}K$,$BB^{*}\pi$, $B^{*}B_{s}K$, $BB_{s}^{*}K$, $DD\rho$, $DD_{s}K^{*}$, $BB_{s}K^{*}$, $DD^{*}\rho$, $DD^{*}_{s}K^{*}$, $BB^{*}_{s}K^{*}$, $D^{*}D^{*}\rho$, $B^{*}B^{*}\rho$, $BB_{s_{0}}K$, $B^{*}B_{s_{1}}K$, $DD^{*}_{s}K_{1}$, $BB^{*}_{s}K_{1}$, $D_{s}^{*}D_{s}\phi$, $D_{2}^{*}D^{*}\pi$, $D_{s_{2}}^{*}D^{*}K$, $D_{2}^{*}D\rho$, $D_{2}^{*}D\omega$, $D_{s_{2}}^{*}D_{s}\phi$, $DDJ/\psi$, $DD^{*}J/\psi$, $D^{*}D^{*}J/\psi$, $DD^{*}\eta_{c}$, $D^{*}D^{*}\eta_{c}$ and $D_{s}D^{*}_{s}\eta_{c}$ have been analyzed with the three-point QCD sum rules (QCDSR)\cite{Navarra:2000ji,Navarra:2001ju,RodriguesdaSilva:2003hh,Bracco:2004rx,Bracco:2006xf,Bracco:2007sg,Bracco:2010bf,OsorioRodrigues:2010fen,Azizi:2010jj,Sundu:2011vz,Cerqueira:2011za,Cui:2011zq,Cui:2012wk,Bracco:2011pg,Yu:2015xwa,Yu:2019sqp,Li:2015xka,Rodrigues:2017qsm,Lu:2023gmd}
, and the coupling constants of vertices $DD^{*}\pi$, $D^{*}D_{s}K$, $DD_{s}^{*}K$,$BB^{*}\pi$, $DD\rho$, $DD_{s}K^{*}$, $D_{s}D_{s}\phi$, $BB\rho$, $DD^{*}\rho$, $D_{s}D^{*}K^{*}$, $D_{s}D^{*}_{s}\phi$, $BB^{*}\rho$, $D^{*}D^{*}\pi$, $D^{*}D^{*}_{s}K$, $B^{*}B^{*}\pi$, $D^{*}D^{*}\rho$, $DD_{0}\pi$, $BB_{0}\pi$, $D_{0}D_{s}K$, $DD_{s_{0}}K$, $BB_{s_{0}}K$, $D_{1}D^{*}\pi$, $B_{1}B^{*}\pi$, $D_{s_{1}}D^{*}K$, $B_{s_{1}}B^{*}K$, $B_{0}B_{1}\pi$, $B_{1}B_{2}\pi$, $B_{2}B^{*}\pi$, $B_{1}B^{*}\rho$, $BB_{1}\rho$, $B_{2}B^{*}\rho$ and $B_{1}B_{2}\rho$ have been studied with the light-cone QCD sum rules (LCSR)\cite{Colangelo:1995ph,Aliev:1996bp,Colangelo:1997rp,Dai:1998ve,Zhu:1998vf,Khodjamirian:1999hb,Li:2002pp,Kim:2001es,Wang:2006bs,Wang:2006ida,Wang:2007mc,Wang:2007zm,Wang:2008tm,Wang:2007ci,Li:2007dv,Khodjamirian:2020mlb}. In the early QCDSR and LCSR calculations, only the leading order contributions are considered, and the higher-order QCD corrections and subleading order contributions are often omitted. In recent years, some strong coupling constants were analyzed with the method of LCSR by considering the higher-order QCD corrections and next leading power contributions\cite{Khodjamirian:2020mlb,Khodjamirian:2011jp}, or with the method of QCDSR by considering higher dimension condensate terms\cite{Yu:2015xwa,Yu:2019sqp,Li:2015xka,Rodrigues:2017qsm,Lu:2023gmd}. These studies indicated that it is important for us to consider the higher-order contributions for the accuracy of the finally results.

In our previous works, the strong vertices $B_{c}B_{c}J/\psi$, $B_{c}B_{c}\Upsilon$, $B_{c}B_{c}^{*}J/\psi$ and $B_{c}B_{c}^{*}\Upsilon$ have been studied by using the three-point QCD sum rules, and the operator product expansion(OPE) was truncated at dimension of 4\cite{Wang:2013iia}. Recently, a systematic analysis of the strong vertices of charmed mesons $D$, $D^{*}$ and charmonia $J/\psi$, $\eta_{c}$ was performed in Ref\cite{Lu:2023gmd}, and the OPE was truncated at dimension 7. As a continuation of these works, the strong vertices of the bottom mesons $B$, $B^{*}$ and the bottomonia $\Upsilon$, $\eta_{b}$ are systematically analyzed in the present work.

This article is organized as follows. After the introduction in Sec. \ref{sec1}, the strong vertices $BB\Upsilon$, $BB^{*}J/\psi$, $B^{*}B^{*}\Upsilon$, $BB^{*}\eta_{b}$ and $B^{*}B^{*}\eta_{b}$ are analyzed with the QCDSR in Sec. \ref{sec2}, in which all off-shell cases of the intermediate mesons are considered. In the QCD side, the perturbative contribution and vacuum condensate terms are considered including $\langle\overline{q}q\rangle$, $\langle\overline{q}g_{s}\sigma Gq\rangle$, $\langle g_{s}^{2}G^{2}\rangle$, $\langle f^{3}G^{3}\rangle$ and $\langle\overline{q}q\rangle\langle g_{s}^{2}G^{2}\rangle$. Sec. \ref{sec3} presents the numerical results and discussions. Sec. \ref{sec4} is reserved for our conclusions.

\section{QCD sum rules}\label{sec2}

To begin with this work, the following three-point correlation function is firstly introduced,
\begin{eqnarray}
\notag
\Pi (p,p') &&= {i^2}\int {{d^4}x} {d^4}y{e^{ip'x}}{e^{i(p - p')y}}\\
&&\times \left\langle0\right.| T\{ {J_{M_3}}(x){J_{M_2}}(y)J_{M_1}^ + (0)\} |\left.0\right\rangle
\end{eqnarray}
where $T$ is the time ordered product, $J$ is the meson interpolating current, the subscripts $M_{1}$, $M_{2}$ and $M_{3}$ denote the mesons in each vertex. Here, $M_{2}$ represents the intermediate meson which is off-shell. The assignments of the mesons for each vertex are shown in Table~\ref{OS}.

\begin{table}[htbp]
\caption{The assignments of the mesons M$_{1}$, M$_{2}$ and M$_{3}$ for each vertex where M$_{2}$ denotes the off-shell mesons.}
\label{OS}
\begin{tabular}{p{1.8cm}<{\centering} p{1.8cm}<{\centering} p{1.8cm}<{\centering} p{1.8cm}<{\centering} }
\hline
\hline
Vertices&M$_{1}$&M$_{2}$(off-shell)&M$_{3}$ \\ \hline
\multirow{2}*{$BB\Upsilon$}&$B$&$\Upsilon$&$B$  \\
~&$B$&$B$&$\Upsilon$   \\  \hline
\multirow{3}*{$BB^{*}\Upsilon$}&$B^{*}$&$\Upsilon$&$B$  \\
~&$B^{*}$&$B$&$\Upsilon$   \\
~&$B$&$B^{*}$&$\Upsilon$   \\ \hline
\multirow{2}*{$B^{*}B^{*}\Upsilon$}&$B^{*}$&$\Upsilon$&$B^{*}$  \\
~&$B^{*}$&$B^{*}$&$\Upsilon$   \\ \hline
\multirow{3}*{$BB^{*}\eta_{b}$}&$B$&$\eta_{b}$&$B^{*}$  \\
~&$B^{*}$&$B$&$\eta_{b}$   \\
~&$B$&$B^{*}$&$\eta_{b}$  \\ \hline
\multirow{2}*{$B^{*}B^{*}\eta_{b}$}&$B^{*}$&$\eta_{b}$&$B^{*}$  \\
~&$B^{*}$&$B^{*}$&$\eta_{b}$ \\ \hline\hline
\end{tabular}
\end{table}

The bottom meson and bottomonium interpolating currents are taken as the following forms,
\begin{eqnarray}
\notag
{J_B}(x) = \bar u(x)i{\gamma _5}b(x)\\
\notag
{J_{{B^*}}}(x) = \bar u(x){\gamma _\mu }b(x)\\
\notag
{J_{\Upsilon }}(x) = \bar b(x){\gamma _\mu }b(x)\\
{J_{{\eta _b}}}(x) = \bar b(x)i{\gamma _5}b(x)
\end{eqnarray}
The correlation function will be calculated at two sides which are called the phenomenological side and the QCD side, respectively. According to the quark hadron duality, calculations of these two sides will be coordinated and the QCD sum rules about the properties of hadrons can be obtained.

\subsection{The Phenomenological side}\label{sec2.1}

In phenomenological side, a complete sets of the hadronic states with the same quantum numbers as the interpolating currents $J_{M_{1}}^{+}$, $J_{M_{2}}$ and $J_{M_{3}}$ are inserted into the correlation function. Then, the correlation function can be written as the following form by using the dispersion relation\cite{Bracco:2011pg},
\begin{eqnarray}\label{eq:3}
\notag
\Pi (p,p') &&= \frac{{\left\langle 0 \right.|{J_{M_3}}(0)|\left. {{M_3}(p')} \right\rangle \left\langle 0 \right.|{J_{M_2}}(0)|\left. {{M_2}(q)} \right\rangle }}{{(m_{M_1}^2 - {p^2})(m_{M_3}^2 - p{'^2})(m_{M_2}^2 - {q^2})}}\\
\notag
&&\times \left\langle {M_1(p)} \right.|J_{M_1}^ + (0)|\left. 0 \right\rangle \left\langle {{M_2}(q){M_3}(p')|\left. {{M_1}(p)} \right\rangle } \right.\\
&&+ ...
\end{eqnarray}
where ellipsis denotes the contributions of higher resonances and continuum states. The meson vacuum matrix elements in Eq. (\ref{eq:3}) are expressed as the following forms,
\begin{eqnarray} \label{eq:4}
\notag
&&\langle0|J_{B}(0)|B\rangle =\frac{f_{B}m_{B}^{2}}{m_{b}}\\
\notag
&&\langle0|J_{B^{*}}(0)|B^{*}\rangle=f_{B^{*}}m_{B^{*}}\zeta _{\mu }\\
\notag
&&\langle0|J_{\Upsilon }(0)|\Upsilon\rangle=f_{\Upsilon}m_{\Upsilon}\xi_{\mu}\\
&&\langle0|J_{\eta _{b}}(0)|\eta _{b}\rangle=\frac{f_{\eta_{b}}m_{\eta_{b}}^{2}}{2m_{b}}
\end{eqnarray}
where $f_{B}$, $f_{B^{*}}$, $f_{\Upsilon}$ and $f_{\eta_{b}}$ are the meson decay constants, $\xi_{\mu}$ and $\zeta_{\mu}$ are the polarization vectors of $\Upsilon$ and $B^{*}$, respectively. All of the meson vertex matrix elements in Eq. (\ref{eq:3}) can be obtained by the following effective Lagrangian,
\begin{eqnarray} \label{eq:5}
\notag
{\mathscr{L}}&&= i{g_{BB\Upsilon }}{\Upsilon _\alpha }({\partial ^\alpha }B\bar B  - B{\partial ^\alpha }\bar B ) - {g_{{B^*}{B^*}{\eta _b}}}{\varepsilon ^{\alpha \beta \rho \tau }}{\partial _\alpha }B_\beta ^*{\partial _\rho }\bar B_\tau ^*{\eta _b} \\
\notag
&& - {g_{{B^*}B\Upsilon }}{\varepsilon ^{\alpha \beta \rho \tau }}{\partial _\alpha }{\Upsilon _\beta }({\partial _\rho }B_\tau ^*\bar B  + B{\partial _\rho }\bar B _\tau ^*) \\
\notag
&&+ i{g_{{B^*}{B^*}\Upsilon }}[{\Upsilon ^\alpha }({\partial _\alpha }{B^{*\beta }}\bar {B_\beta ^*}  - {B^{*\beta }}{\partial _\alpha }\bar {B_\beta ^*} ) \\
\notag
&&+ ({\partial _\alpha }{\Upsilon _\beta }{B^{*\beta }} - {\Upsilon _\beta }{\partial _\alpha }{B^{*\beta }}){\bar B ^{*\alpha }} + {B^{*\alpha }}({\Upsilon ^\beta }{\partial _\alpha }\bar {B_\beta ^*}  - {\partial _\alpha }{\Upsilon _\beta }{\bar B ^{*\beta }})]  \\
\notag
&&+ i{g_{{B^*}B{\eta _b}}}[{B^{*\alpha }}({\partial _\alpha }{\eta _b}\bar B  - {\eta _b}{\partial _\alpha }\bar B ) \\
&&+ ({\partial _\alpha }{\eta _b}B - {\eta _b}{\partial _\alpha }B){\bar B ^{*\alpha }}]
\end{eqnarray}
According to this Lagrangian, all of the vertex matrix elements can be written as,
\begin{eqnarray} \label{eq:6}
\notag
&&\left\langle {B(p')\Upsilon (q)|\left. {B(p)} \right\rangle } \right. = g_{BB\Upsilon }^{\Upsilon }({q^2})\xi _\alpha ^*{(p + p')^\alpha }\\
\notag
&&\left\langle {B(q)\Upsilon (p')|\left. {B(p)} \right\rangle } \right. = g_{BB\Upsilon }^B({q^2}){\xi _\alpha }{(p + q)^\alpha }\\
\notag
&&\left\langle {B(p')\Upsilon (q)|\left. {{B^*}(p)} \right\rangle } \right. =  - g_{B{B^*}\Upsilon }^{\Upsilon }({q^2}){\varepsilon ^{\alpha \beta \rho \tau }}{\xi _\alpha }{\zeta _\beta }{p_\rho }p{'_\tau }\\
\notag
&&\left\langle {B(q)\Upsilon (p')|\left. {{B^*}(p)} \right\rangle } \right. =  - g_{B{B^*}\Upsilon }^B({q^2}){\varepsilon ^{\alpha \beta \rho \tau}}{\xi _\alpha }{\zeta _\beta}{p_\rho }p{'_\tau }\\
\notag
&&\left\langle {{B^*}(q)\Upsilon (p')|\left. {B(p)} \right\rangle } \right. =  g_{B{B^*}\Upsilon }^{{B^*}}({q^2}){\varepsilon ^{\alpha \beta \rho \tau}}{\xi _\alpha}{\zeta _\beta}p{'_\rho}{p_\tau }\\
\notag
&&\left\langle {{B^*}(p')\Upsilon (q)|\left. {{B^*}(p)} \right\rangle } \right. = g_{{B^*}{B^*}J/\psi }^{\Upsilon }[({p^\alpha } + p{'^\alpha }){\xi^{*}_{\alpha }}{\zeta^{'\beta} }\zeta _\beta ^*  \\
\notag
&&- ({p^\alpha } + {q^\alpha }){\zeta^{'*}_{\alpha }\xi _\beta ^*{\zeta ^\beta }} - (p{'^\alpha } - {q^\alpha }){\zeta _\alpha }{\xi ^{*\beta }}\zeta _{\beta}^{*}]\\
\notag
&&\left\langle {{B^*}(q)\Upsilon (p')|\left. {{B^*}(p)} \right\rangle } \right. = g_{{B^*}{B^*}\Upsilon }^{{B^*}}[({p^\alpha } + {q^\alpha })\xi^{*}_{\alpha}{\zeta^{'\beta}}\varepsilon _\beta ^* \\
\notag
&&- ({p^\alpha } + p{'^\alpha }){\zeta^{'*}_{\alpha }} \xi _\beta ^*{\zeta ^\beta }- ({q^\alpha } - p{'^\alpha }){\zeta _\alpha }{\xi ^{*\beta }}\zeta _{\beta}^{'*}]\\
\notag
&&\left\langle {B(p'){\eta _b}(q)|\left. {{B^*}(p)} \right\rangle } \right. =  - g_{B{B^*}{\eta _b}}^{{\eta _b}}({q^2}){\zeta _\alpha }(q-p')^{\alpha}\\
\notag
&&\left\langle {B(q){\eta _b}(p')|\left. {{B^*}(p)} \right\rangle } \right. =  - g_{B{B^*}{\eta _b}}^B({q^2}){\zeta _\alpha }(p'-q)^{\alpha}\\
\notag
&&\left\langle {{B^*}(q){\eta _b}(p')|\left. {B(p)} \right\rangle } \right. =  - g_{B{B^*}{\eta _b}}^{{B^*}}({q^2})\zeta^{*}_{\alpha}(p+p')^{\alpha}\\
\notag
&&\left\langle {{B^*}(p'){\eta _b}(q)|\left. {{B^*}(p)} \right\rangle } \right. =  - g_{{B^*}{B^*}{\eta _b}}^{{\eta _b}}({q^2}){\varepsilon ^{\alpha \beta \rho \tau }}\zeta _\alpha \zeta^{'*} _{\beta}{p_\rho}p{'_\tau }\\
&&\left\langle {{B^*}(q){\eta _b}(p')|\left. {{B^*}(p)} \right\rangle } \right. =  - g_{{B^*}{B^*}{\eta _b}}^{{D^*}}({q^2}){\varepsilon ^{\alpha \beta \rho \tau }}\zeta^{'}_{\alpha} \zeta^{*}_\beta {p_\rho}{q_\tau }
\end{eqnarray}
where $\xi_{\alpha}$ and $\zeta^{(')}_{\alpha}$ are the polarization vectors of $\Upsilon$ and $B^{*}$ respectively, $q=p-p'$, and $\varepsilon^{\alpha\beta\rho\tau}$ is the 4-dimension Levi-Civita tensor. The subscript of $g$ in Eq. (\ref{eq:6}) denotes the type of strong vertex, and the superscript denotes the intermediate meson which is off-shell. From Eqs. (\ref{eq:3}) $\sim$ (\ref{eq:6}), the expressions of the correlation function in phenomenological side can be obtained, and can be divided into different tensor structures. In general, different tensor structures for a correlation function will lead to the same result, thus choosing an appropriate structure to analyze the strong vertex is an acceptable way.

\subsection{The QCD side}\label{sec2.2}
In QCD side, we will contract the quark fields with Wick's theorem and then do the operator product expansion(OPE). After the first process, the correlation functions for vertices $BB\Upsilon$, $BB^{*}\Upsilon$, $B^{*}B^{*}\Upsilon$, $BB^{*}\eta_{b}$ and $B^{*}B^{*}\eta_{b}$ can be written as,

\begin{eqnarray}\label{eq:7}
\notag
\Pi _\mu ^{\Upsilon }(p,p') &&= \int {{d^4}x{d^4}y{e^{ip'x}}{e^{i(p - p')y}}} \\
\notag
&&\times Tr\{ {B^{nk}}(y){\gamma _5}{U^{km}}( - x){\gamma _5}{B^{mn}}(x - y){\gamma _\mu }\} \\
\notag
\Pi _\mu ^B(p,p') &&= \int {{d^4}x{d^4}y{e^{ip'x}}{e^{i(p - p')y}}} \\
&&\times Tr\{ {\gamma _\mu }{B^{nk}}(x){\gamma _5}{U^{km}}( - y){\gamma _5}{B^{mn}}(y - x)\}
\end{eqnarray}
\begin{eqnarray}\label{eq:8}
\notag
\Pi _{\mu \nu }^{\Upsilon }(p,p') &&=  - i\int {{d^4}x{d^4}y{e^{ip'x}}{e^{i(p - p')y}}}\\
\notag
&&\times Tr\{ {B^{nk}}(y){\gamma _\nu }{U^{km}}( - x){\gamma _5}{B^{mn}}(x - y){\gamma _\mu }\} \\
\notag
\Pi _{\mu \nu }^B(p,p') &&=  - i\int {{d^4}x{d^4}y{e^{ip'x}}{e^{i(p - p')y}}} \\
\notag
&&\times Tr\{ {\gamma _\mu }{B^{nk}}(x){\gamma _\nu }{U^{km}}( - y){\gamma _5}{B^{mn}}(y - x)\}  \\
\notag
\Pi _{\mu \nu }^{{B^*}}(p,p') &&=  - i\int {{d^4}x{d^4}y{e^{ip'x}}{e^{i(p - p')y}}} \\
&&\times Tr\{ {\gamma _\mu }{B^{nk}}(x){\gamma _5}{U^{km}}( - y){\gamma _\nu }{B^{mn}}(y - x)\}
\end{eqnarray}
\begin{eqnarray}\label{eq:9}
\notag
\Pi _{\mu \nu \sigma }^{\Upsilon }(p,p') &&= \int {{d^4}x{d^4}y{e^{ip'x}}{e^{i(p - p')y}}}  \\
\notag
&&\times Tr\{ {B^{nk}}(y){\gamma _\nu }{U^{km}}( - x){\gamma _\sigma }{B^{mn}}(x - y){\gamma _\mu } \} \\
\notag
\Pi _{\mu \nu \sigma }^{{B^*}}(p,p') &&= \int {{d^4}x{d^4}y{e^{ip'x}}{e^{i(p - p')y}}} \\
&&\times Tr\{ {\gamma _\mu }{B^{nk}}(x){\gamma _\sigma }{U^{km}}( - y){\gamma _\nu }{B^{mn}}(y - x)\}
\end{eqnarray}
\begin{eqnarray}\label{eq:10}
\notag
\Pi _\mu ^{{\eta _b}}(p,p') &&= \int {{d^4}x{d^4}y{e^{ip'x}}{e^{i(p - p')y}}} \\
\notag
&&\times Tr\{ {B^{nk}}(y){\gamma _\mu }{U^{km}}( - x){\gamma _5}{B^{mn}}(x - y){\gamma _5}\} \\
\notag
\Pi _\mu ^B(p,p') &&= \int {{d^4}x{d^4}y{e^{ip'x}}{e^{i(p - p')y}}} \\
\notag
&&\times Tr\{ {\gamma _5}{B^{nk}}(x){\gamma _\mu }{U^{km}}( - y){\gamma _5}{B^{mn}}(y - x)\}  \\
\notag
\Pi _\mu ^{{B^*}}(p,p') &&= \int {{d^4}x{d^4}y{e^{ip'x}}{e^{i(p - p')y}}}  \\
&&\times Tr\{ {\gamma _5}{B^{nk}}(x){\gamma _5}{U^{km}}( - y){\gamma _\mu }{B^{mn}}(y - x)\}
\end{eqnarray}
\begin{eqnarray}\label{eq:11}
\notag
\Pi _{\mu \nu }^{{\eta _b}}(p,p') &&=  - i\int {{d^4}x{d^4}y{e^{ip'x}}{e^{i(p - p')y}}} \\
\notag
&&\times Tr\{ {B^{nk}}(y){\gamma _\nu }{U^{km}}( - x){\gamma _\mu }{B^{mn}}(x - y){\gamma _5}\}  \\
\notag
\Pi _{\mu \nu }^{{B^*}}(p,p') &&=  - i\int {{d^4}x{d^4}y{e^{ip'x}}{e^{i(p - p')y}}} \\
&&\times Tr\{ {\gamma _5}{B^{nk}}(x){\gamma _\mu }{U^{km}}( - y){\gamma _\nu }{B^{mn}}(y - x)\}
\end{eqnarray}
The superscripts of $\Pi$ in these above equations denote the intermediate mesons. $U^{ij}(x)$ and $B^{ij}(x)$ are the full propagators of $u(d)$ and $b$ quarks which have the following forms\cite{Reinders:1984sr},
\begin{eqnarray}
\notag
{U^{ij}}(x) &&= \frac{{i{\delta ^{ij}}x\!\!\!/}}{{2{\pi ^2}{x^4}}} - \frac{{{\delta ^{ij}}{m_q}}}{{4{\pi ^2}{x^4}}} - \frac{{{\delta ^{ij}}\left\langle {\bar qq} \right\rangle }}{{12}} + \frac{{i{\delta ^{ij}}x\!\!\!/{m_q}\left\langle {\bar qq} \right\rangle }}{{48}}\\
\notag
&& - \frac{{{\delta ^{ij}}{x^2}\left\langle {\bar q{g_s}\sigma Gq} \right\rangle }}{{192}} + \frac{{i{\delta ^{ij}}{x^2}x\!\!\!/{m_q}\left\langle {\bar q{g_s}\sigma Gq} \right\rangle }}{{1152}}\\
\notag
&& - \frac{{i{g_s}G_{\alpha \beta }^at_{ij}^a(x\!\!\!/{\sigma ^{\alpha \beta }} + {\sigma ^{\alpha \beta }}x\!\!\!/)}}{{32{\pi ^2}{x^2}}} - \frac{{i{\delta ^{ij}}{x^2}x\!\!\!/g_s^2{{\left\langle {\bar qq} \right\rangle }^2}}}{{7776}}\\
\notag
&& - \frac{{{\delta ^{ij}}{x^4}\left\langle {\bar qq} \right\rangle \left\langle {g_s^2GG} \right\rangle }}{{27648}} - \frac{{\left\langle {{{\bar q}^j}{\sigma ^{\mu \nu }}{q^i}} \right\rangle {\sigma _{\mu \nu }}}}{8}\\
\notag
&& - \frac{{\left\langle {{{\bar q}^j}{\gamma ^\mu }{q^i}} \right\rangle {\gamma _\mu }}}{4} + ...\\
\notag
{B^{ij}}(x) &&= \frac{i}{{{{(2\pi )}^4}}}\int {{d^4}k} {e^{ - ik \cdot x}}\{ \frac{{{\delta ^{ij}}}}{{k\!\!\!/ - {m_b}}}\\
\notag
&& - \frac{{{g_s}G_{\alpha \beta }^nt_{ij}^n}}{4}\frac{{{\sigma ^{\alpha \beta }}(k\!\!\!/ + {m_b}) + (k\!\!\!/ + {m_b}){\sigma ^{\alpha \beta }}}}{{{{({k^2} - m_b^2)}^2}}}\\
\notag
&& + \frac{{{g_s}{D_\alpha }G_{\beta \lambda }^nt_{ij}^n({f^{\lambda \beta \alpha }} + {f^{\lambda \alpha \beta }})}}{{3{{({k^2} - m_b^2)}^4}}}\\
\notag
&& - \frac{{g_s^2{{({t^a}{t^b})}_{ij}}G_{\alpha \beta }^aG_{\mu \nu }^b({f^{\alpha \beta \mu \nu }} + {f^{\alpha \mu \beta \nu }} + {f^{\alpha \mu \nu \beta }})}}{{4{{({k^2} - m_b^2)}^5}}} \\
&&+ ...\}
\end{eqnarray}
where $\langle g_{s}^{2}G^{2}\rangle=\langle g_{s}^{2}G^{n}_{\alpha\beta}G^{n\alpha\beta}\rangle$, $D_{\alpha}=\partial_{\alpha}-ig_{s}G^{n}_{\alpha}t^{n}$, $t^{n}=\frac{\lambda^{n}}{2}$. $\lambda^{n}(n=1,...,8)$ are the Gell-Mann matrixs, $i$ and $j$ are color indices, $q=u(d)$, $\sigma_{\alpha\beta}=\frac{i}{2}[\gamma_{\alpha},\gamma_{\beta}]$, $f^{\lambda\alpha\beta}$ and $f^{\alpha\beta\mu\nu}$ have the following forms,
\begin{eqnarray}
{f^{\lambda \alpha \beta }} = (k\!\!\!/ + {m_b}){\gamma ^\lambda }(k\!\!\!/ + {m_b}){\gamma ^\alpha }(k\!\!\!/ + {m_b}){\gamma ^\beta }(k\!\!\!/ + {m_b})
\end{eqnarray}
\begin{eqnarray}
\notag
{f^{\alpha \beta \mu \nu }} = && (k\!\!\!/ + {m_b}){\gamma ^\alpha }(k\!\!\!/ + {m_b}){\gamma ^\beta }(k\!\!\!/ + {m_b})\\
&&{\gamma ^\mu }(k\!\!\!/ + {m_b}){\gamma ^\nu }(k\!\!\!/ + {m_b})
\end{eqnarray}
Just as stated in Sec. \ref{sec2.1}, different correlation functions $\Pi_{\mu}$, $\Pi_{\mu\nu}$ and $\Pi_{\mu\nu\sigma}$ in Eqs. (\ref{eq:7}) $\sim$ (\ref{eq:11}) can be expanded into different tensor structures,
\begin{eqnarray}
\notag
&&\Pi_{\mu}(p,p')=\Pi_{1}(p^{2},p'^{2},q^{2})p_{\mu}+ \Pi_{2}(p^{2},p'^{2},q^{2})p'_{\mu} \\
\notag
&&\Pi_{\mu\nu}(p,p')= \Pi(p^{2},p'^{2},q^{2})\varepsilon_{\mu\nu\alpha\beta}p^{\alpha}p'^{\beta} \\
\notag
&&{\Pi _{\mu \nu \sigma }}(p,p') = {\Pi _1}({p^2},p{'^2},{q^2}){p_\mu }{g_{\nu \sigma }} \\
\notag
&& + {\Pi _2}({p^2},p{'^2},{q^2}){p_\mu }{p_\nu }{p_\sigma } + {\Pi _3}({p^2},p{'^2},{q^2})p{'_\mu }{p_\nu }{p_\sigma }\\
\notag
&& + {\Pi _4}({p^2},p{'^2},{q^2}){p_\sigma }{g_{\mu \nu }} + {\Pi _5}({p^2},p{'^2},{q^2}){p_\mu }p{'_\nu }{p_\sigma }\\
\notag
&& + {\Pi _6}({p^2},p{'^2},{q^2})p{'_\mu }{g_{\nu \sigma }} + {\Pi _7}({p^2},p{'^2},{q^2}){p_\mu }{p_\nu }p{'_\sigma }\\
\notag
&& + {\Pi _8}({p^2},p{'^2},{q^2})p{'_\nu }{g_{\mu \sigma }} + {\Pi _{9}}({p^2},p{'^2},{q^2})p{'_\mu }p{'_\nu }{p_\sigma }\\
\notag
&& + {\Pi _{10}}({p^2},p{'^2},{q^2})p{'_\sigma }{g_{\mu \nu }} + {\Pi _{11}}({p^2},p{'^2},{q^2})p{'_\mu }{p_\nu }p{'_\sigma }\\
\notag
&& + {\Pi _{12}}({p^2},p{'^2},{q^2}){p_\mu }p{'_\nu }p{'_\sigma }+ {\Pi _{13}}({p^2},p{'^2},{q^2}){p_\nu }{g_{\mu \sigma }}\\
&&+ {\Pi _{14}}({p^2},p{'^2},{q^2})p{'_\mu }p{'_\nu }p{'_\sigma }
\end{eqnarray}
where $g_{\mu\nu}$ is the metric tensor. In the right side of these above equations, $\Pi$ without Lorentz index is commonly called scalar invariant amplitude. To obtain the strong coupling constant, an appropriate scalar amplitude should be selected to carry out the analysis\cite{Bracco:2004rx}. For vertex $B^{*}B^{*}\Upsilon$ as an example, its correlation function $\Pi_{\mu\nu\sigma}$ has fourteen tensor structures. In principle, it is reasonable to perform the calculations with each structure, in this paper we will choose the structure $p_{\mu}g_{\nu\sigma}$ to analyze the strong vertex $B^{*}B^{*}\Upsilon$.

\begin{figure}[htbp]
\centering
\includegraphics[width=9cm]{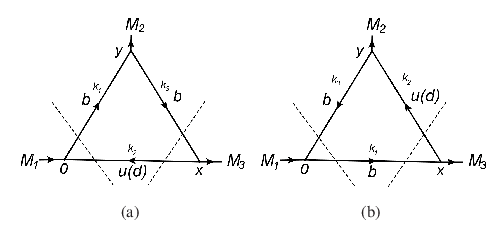}
\caption{The perturbative contributions for $\Upsilon(\eta_{b})$ (a) and $B(B^{*})$ (b) off-shell. The dashed lines denote the Cutkosky cuts.}
\label{fig:free}
\end{figure}

\begin{figure*}[htbp]
\centering
\includegraphics[width=18cm]{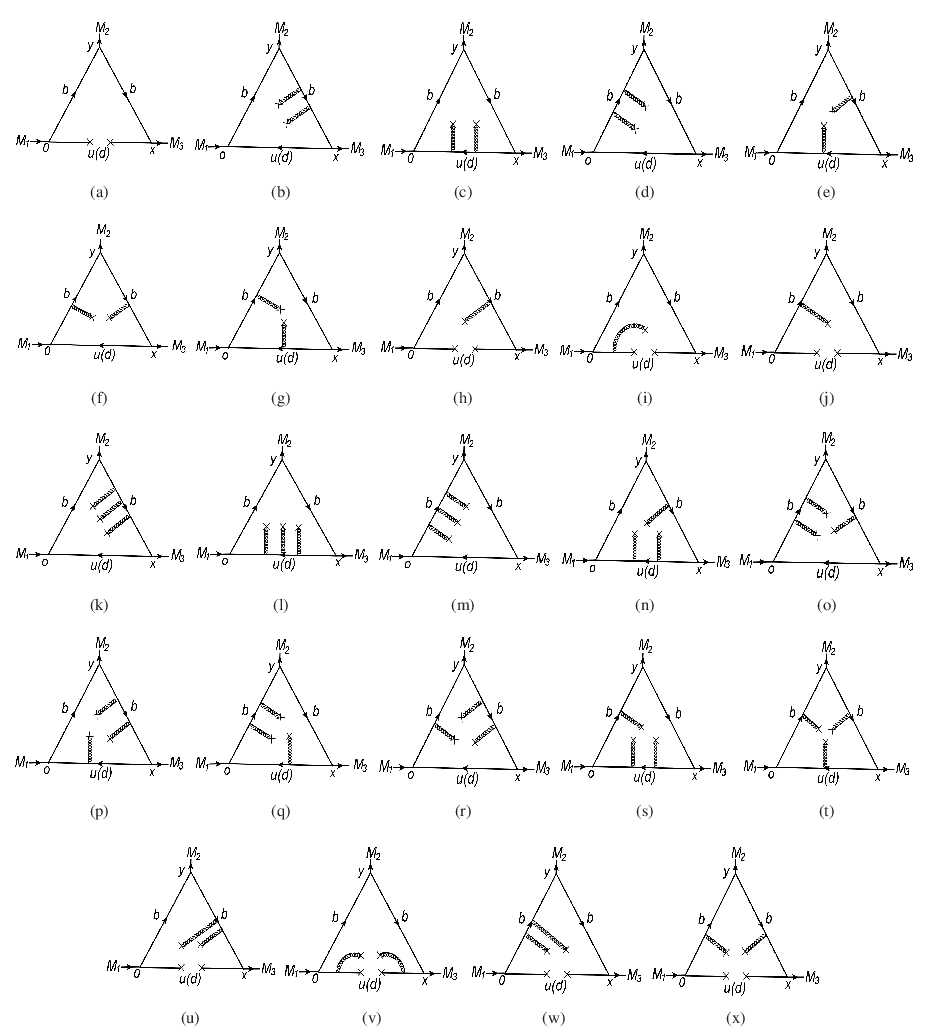}
\caption{Contributions of the non-perturbative parts for $\Upsilon(\eta_{b})$ off-shell.}
\label{fig:FM1}
\end{figure*}

\begin{figure*}[htbp]
\centering
\includegraphics[width=16cm]{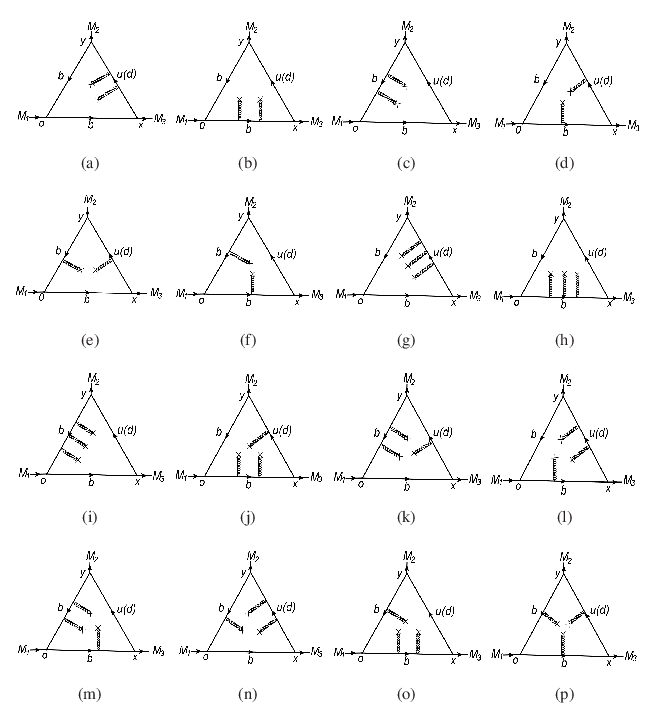}
\caption{Contributions of the non-perturbative parts for $B(B^{*})$ off-shell.}
\label{fig:FM2}
\end{figure*}

The scalar invariant amplitudes in the QCD side are represented as $\Pi^{\mathrm{OPE}}$ which can be divided into two parts,
\begin{eqnarray}
\Pi^{\mathrm{OPE}}=\Pi^{\mathrm{pert}}+\Pi^{\mathrm{non-pert}}
\end{eqnarray}
where $\Pi^{\mathrm{pert}}$ refers to the perturbative part and $\Pi^{\mathrm{non-pert}}$ denotes the non-perturbative contributions including $\langle\bar{q}q\rangle$, $\langle g_{s}^{2}G^{2}\rangle$, $\langle\bar{q}g_{s}\sigma Gq\rangle$, $\langle f^{3}G^{3}\rangle$ and $\langle\bar{q}q\rangle\langle g_{s}^{2}G^{2}\rangle$. The perturbative part, $\langle g_{s}^{2}G^{2}\rangle$ and $\langle f^{3}G^{3}\rangle$ terms can be written as the following form according to the dispersion relation,
\begin{eqnarray}
\notag
\Pi }(p,p') =  - \int\limits_0^\infty  {\int\limits_0^\infty  {\frac{{{\rho }(s,u,{q^2})}}{{(s - {p^2})(u - p{'^2})}}dsdu}
\end{eqnarray}
where
\begin{eqnarray}
\notag
\rho (s,u,{q^2}) &&= {\rho ^{\mathrm{pert}}}(s,u,{q^2}) + {\rho ^{\left\langle {g_s^2{G^2}} \right\rangle }}(s,u,{q^2}) \\
&&+ {\rho ^{\left\langle {{f^3}{G^3}} \right\rangle }}(s,u,{q^2})
\end{eqnarray}
and $s=p^{2}$, $u=p'^{2}$ and $q=p-p'$. The QCD spectral density $\rho(s,u,q^{2})$ can be obtained by the Cutkosky's rules\cite{Wang:2007ys}(see Fig. \ref{fig:free}), where the calculation detail has already been discussed in Ref. \cite{Lu:2023gmd}. Besides, we also take into account the contributions of $\langle\overline{q}q\rangle$, $\langle\overline{q}g_{s}\sigma Gq\rangle$, and $\langle\overline{q}q\rangle\langle g_{s}^{2}G^{2}\rangle$. The feynman diagrams for all of these condensate terms can be classified into two groups which are illustrated in Figs. \ref{fig:FM1} and \ref{fig:FM2}.

We take the change of variables $p^{2}\to-P^{2}$, $p'^{2}\to-P'^{2}$ and $q^{2}\to-Q^{2}$ and perform double Borel transformation to both the phenomenological and QCD sides. The variables $P^{2}$ and $P'^{2}$ are replaced by $T_{1}^{2}$ and $T_{2}^{2}$, where $T_{1}$ and $T_{2}$ are called the Borel parameters. In this article, we take $T^{2}=T_{1}^{2}$ and $T_{2}^{2}=kT_{1}^{2}=kT^{2}$, where $k$ is a constant which is related to meson mass. It takes different values for different vertices, where these values are represented in Table \ref{kE}. Finally, the sum rules for the coupling constants can be obtained by matching the phenomenological and QCD sides according to the quark-hadron duality. The momentum dependent strong coupling constant can be written as,
\begin{align}\label{eq:20}
g({Q^2}) = \frac{{ - \int\limits_{{s_1}}^{{s_0}} {\int\limits_{{u_1}}^{{u_0}} {\rho (s,u,{Q^2}){e^{ - s/{T^2}}}{e^{ - u/k{T^2}}}} dsdu + \mathscr B\mathscr B[{{\Pi }^{\mathrm{non - pert}}}]} }}{{\frac{E}{{(m_{M_2}^2 + {Q^2})}}{e^{ - m_{M_1}^2/{T^2}}}{e^{ - m_{M_3}^2/k{T^2}}}}}
\end{align}
where $\mathscr{BB}$ stands for the double Borel transformation and $E$ is a factor which is related with meson masses and decay constants(see Table \ref{kE}). From this equation, we can also see that the threshold parameters $s_{0}$ and $u_{0}$ in dispersion integral will be introduced. These parameters are used to eliminate the contributions of higher resonances and continuum states in Eq. (\ref{eq:3}). They commonly fulfill the relations, $m_{\mathrm{i}}^{2}<s_{0}<m'^{2}_{\mathrm{i}}$ and $m_{\mathrm{o}}^{2}<u_{0}<m'^{2}_{\mathrm{o}}$, where subscripts $i$ and $o$ represent incoming and outcoming mesons, respectively. $m$ and $m'$ are the masses of the ground and first excited state of the mesons. There usually have a relation $m'=m+\Delta$, where $\Delta$ is taken as a value of $0.4 \sim 0.6$ GeV\cite{Bracco:2011pg}.
\begin{table}[htbp]
\caption{The parameters $k$ and $E$ for different vertices and off-shell cases.}
\label{kE}
\begin{tabular}{c c c c}
\hline\hline
Vertices&off-shell&$k$&$E$ \\ \hline
\specialrule{0em}{1pt}{1pt}
\multirow{2}*{$BB\Upsilon$}&$\Upsilon$ &$1$&$\frac{f_{B}^{2}m_{B}^{4}f_{\Upsilon}m_{\Upsilon}}{m_{b}^{2}}$  \\
\specialrule{0em}{1pt}{1pt}
~&$B$ &$\frac{m_{\Upsilon}^{2}}{m_{B^{*}}^{2}}$&$\frac{2f_{B}^{2}m_{B}^{4}f_{\Upsilon}m_{\Upsilon}}{m_{b}^{2}}$\\
\specialrule{0em}{1pt}{1pt} \hline
\specialrule{0em}{1pt}{1pt}
\multirow{3}*{$BB^{*}\Upsilon$}&$\Upsilon$ &$\frac{m_{B^{*}}^{2}}{m_{B}^{2}}$&\multirow{3}*{$-\frac{f_{B}m_{B}^{2}f_{B^{*}}m_{B^{*}}f_{\Upsilon}m_{\Upsilon}}{m_{b}}$}  \\
\specialrule{0em}{1pt}{1pt}
~&$B$ &$\frac{m_{\Upsilon}^{2}}{m_{B^{*}}^{2}}$&~   \\
\specialrule{0em}{1pt}{1pt}
~&$B^{*}$ &$\frac{m_{\Upsilon}^{2}}{m_{B}^{2}}$&~   \\
\specialrule{0em}{1pt}{1pt} \hline
\specialrule{0em}{1pt}{1pt}
\multirow{2}*{$B^{*}B^{*}\Upsilon$}&$\Upsilon$ &$1$&$-f_{B^{*}}^{2}m_{B^{*}}^{2}f_{\Upsilon}m_{\Upsilon}$  \\
\specialrule{0em}{1pt}{1pt}
~&$B^{*}$ &$\frac{m_{\Upsilon}^{2}}{m_{B}^{2}}$&$-2f_{B^{*}}^{2}m_{B^{*}}^{2}f_{\Upsilon}m_{\Upsilon}$  \\
\specialrule{0em}{1pt}{1pt} \hline
\specialrule{0em}{1pt}{1pt}
\multirow{3}*{$BB^{*}\eta_{b}$}&$\eta_{b}$ &$\frac{m_{B^{*}}^{2}}{m_{B}^{2}}$&\multirow{2}*{$-\frac{f_{B}m_{B}^{2}f_{B^{*}}m_{B^{*}}f_{\eta_{b}}m_{\eta_{b}}^{2}}{m_{b}^{2}}$}  \\
\specialrule{0em}{1pt}{1pt}
~&$B$ &$\frac{m_{\eta_{b}}^{2}}{m_{B^{*}}^{2}}$&~   \\
\specialrule{0em}{1pt}{1pt}
~&$B^{*}$ &$\frac{m_{\eta_{b}}^{2}}{m_{B}^{2}}$&$\frac{f_{B}m_{B}^{2}f_{B^{*}}m_{B^{*}}f_{\eta_{b}}m_{\eta_{b}}^{2}(m_{\eta_{b}}^{2}+m_{B^{*}}^{2}-m_{B}^{2})}{2m_{b}^{2}m_{B^{*}}^{2}}$   \\
\specialrule{0em}{1pt}{1pt} \hline
\specialrule{0em}{1pt}{1pt}
\multirow{2}*{$B^{*}B^{*}\eta_{b}$}&$\eta_{b}$ &$1$&\multirow{2}*{$-\frac{f_{B^{*}}^{2}m_{B^{*}}^{2}f_{\eta_{b}}m_{\eta_{b}}^{2}}{2m_{b}^{2}}$}  \\
\specialrule{0em}{1pt}{1pt}
~&$B^{*}$ &$\frac{m_{\eta_{b}}^{2}}{m_{B^{*}}^{2}}$&~   \\
\specialrule{0em}{1pt}{1pt} \hline\hline
\end{tabular}
\end{table}

The strong coupling constant in Eq. (\ref{eq:20}) is momentum dependence, which is similar with the run coupling constant $\alpha_{s}$.
In order to obtain the final results of strong coupling constant, it is necessary to extrapolate the results which are obtained from Eq. (\ref{eq:20}) into time-like regions $(Q^{2}<0)$. This process is realized by fitting the $g(Q^{2})$ into appropriate analytical functions and setting ($Q^2=-m_{\mathrm{on-shell}}^{2}$).

\section{Numerical results and Discussions}\label{sec3}

All of the the input parameters are taken as the standard values such as the hadronic masses and decay constants $m_{B}=5.28$ GeV\cite{ParticleDataGroup:2022pth}, $m_{B^{*}}=5.33$ GeV\cite{ParticleDataGroup:2022pth}, $m_{\Upsilon}=9.46$ GeV\cite{ParticleDataGroup:2022pth}, $m_{\eta_{b}}=9.399$ GeV\cite{ParticleDataGroup:2022pth}, $f_{B}=0.192\pm0.013$ GeV\cite{Wang:2015mxa}, $f_{B^{*}}=0.213\pm0.018$ GeV\cite{Wang:2015mxa}, $f_{\Upsilon}=0.7$ GeV\cite{ParticleDataGroup:2022pth}, $f_{\eta_{b}}=0.667\pm0.007$ GeV\cite{Becirevic:2017chd}. The vacuum condensates are also adopted as the standard values which are $\langle\overline{q}q\rangle=-(0.23\pm0.01)^{3}$ GeV$^{3}$\cite{ParticleDataGroup:2022pth}, $\langle\overline{q}g_{s}\sigma Gq\rangle=m_{0}^{2}\langle\overline{q}q\rangle$\cite{ParticleDataGroup:2022pth}, $m_{0}^{2}=0.8\pm0.1$ GeV$^2$\cite{Narison:2010cg,Narison:2011xe,Narison:2011rn}, $\langle g_{s}^{2}G^{2}\rangle=0.88\pm0.15$ GeV$^{4}$\cite{Narison:2010cg,Narison:2011xe,Narison:2011rn}, $\langle f^{3}G^{3}\rangle=(8.8\pm5.5)$ GeV$^{2}\langle g_{s}^{2}G^{2}\rangle$\cite{Narison:2010cg,Narison:2011xe,Narison:2011rn}. The $\mathrm{\overline{MS}}$ mass of light and heavy quarks is adopted from the Particle Data Group\cite{ParticleDataGroup:2022pth} where $m_{u(d)}(\mu=1 \mathrm{GeV})=0.006\pm0.001$ GeV and $m_{b}(m_{b})=4.18\pm0.03$ GeV. In principle, both the vacuum condensates and the $\mathrm{\overline{MS}}$ masses of light and heavy quarks are energy-scale dependent. For the strong coupling constants of $B$ meson, the influences of vacuum condensates and the masses of light quarks are smaller than that of bottom quark. That is to say, the mass of bottom quark has more significant influence on the results and its energy-scale dependency should be considered and discussed. The energy-scale dependent $\mathrm{\overline{MS}}$ mass of bottom quark can be expressed as the following form according to the re-normalization group equation,
\begin{eqnarray}
\notag
{m_b}(\mu ) &&= {m_b}({m_b}){[\frac{{{\alpha _s}(\mu )}}{{{\alpha _s}({m_b})}}]^{\frac{{12}}{{33 - 2{n_f}}}}}\\
\notag
{\alpha _s}(\mu ) &&= \frac{1}{{{b_0}t}}[1 - \frac{{{b_1}}}{{b_0^2}}\frac{{\log t}}{t} \\
&&+ \frac{{b_1^2({{\log }^2}t - \log t - 1) + {b_0}{b_2}}}{{b_0^4{t^2}}}]
\end{eqnarray}
where $t=\mathrm{log}\frac{\mu^2}{\Lambda_{QCD}^2}$, $b_{0}=\frac{33-2n_{f}}{12\pi}$, $b_{1}=\frac{153-19n_{f}}{24\pi^{2}}$, $b_{2}=\frac{2857-\frac{5033}{9}n_{f}+\frac{325}{27}n_{f}^{2}}{128\pi^{3}}$, $\Lambda_{QCD}=210$ MeV, 292 MeV and 332 MeV for the flavors $n_{f}=5, 4$ and 3, respectively\cite{ParticleDataGroup:2022pth}. In the framework of QCDSR, the final results should be obtained by selecting an appropriate energy-scale. It was indicated by some similar studies that the masses of heavy quarks in energy-scale $\mu=m_{Q}$ are employed to obtain the coupling constants\cite{Bracco:2011pg,Rodrigues:2017qsm}. In the present work, the mass of bottom quark in the energy-scale $\mu=m_{b}$ is also used to obtain the strong coupling constants. The dependence of the strong coupling constants on energy-scales will be discussed in the end of this section.

According to Eq. (\ref{eq:20}), the strong coupling constant is a function of some input parameters such as the Borel parameter $T^{2}$, the continuum thresholds $s_{0}$ and $u_{0}$, and the square momentum $Q^{2}$. The Borel parameter $T^{2}$ is determined by two conditions, which are the pole dominance and convergence of OPE. To analyze the pole contribution, we firstly write down,
\begin{eqnarray}
\notag
\Pi^{\mathrm{OPE}}_{\mathrm{pole}}(T^{2})=-\int_{s_{1}}^{s_{0}}\int_{u_{1}}^{u_{0}}\rho^{\mathrm{OPE}}(s,u,Q^2)e^{-\frac{s}{T^{2}}}e^{-\frac{u}{kT^{2}}}dsdu \\
\Pi^{\mathrm{OPE}}_{\mathrm{cont}}(T^{2})=-\int_{s_{0}}^{\infty}\int_{u_{0}}^{\infty}\rho^{\mathrm{OPE}}(s,u,Q^2)e^{-\frac{s}{T^{2}}}e^{-\frac{u}{kT^{2}}}dsdu
\end{eqnarray}
As for the pole and continuum contributions, they can be defined as\cite{Bracco:2011pg},
\begin{eqnarray}
\notag
\mathrm{Pole}=\frac{\Pi^{\mathrm{OPE}}_{\mathrm{pole}}(T^{2})}{\Pi^{\mathrm{OPE}}_{\mathrm{pole}}(T^{2})+\Pi^{\mathrm{OPE}}_{\mathrm{cont}}(T^{2})} \\
\mathrm{Continuum}=\frac{\Pi^{\mathrm{OPE}}_{\mathrm{cont}}(T^{2})}{\Pi^{\mathrm{OPE}}_{\mathrm{pole}}(T^{2})+\Pi^{\mathrm{OPE}}_{\mathrm{cont}}(T^{2})}
\end{eqnarray}
In addition, the continuum threshold parameters $s_{0}$ and $u_{0}$ are employed to include the pole contribution and suppress the contribution of higher resonances and continuum states. Their values are commonly satisfy the relations  $s_{0}=(m_{\mathrm{i}}+\Delta_{\mathrm{i}})^{2}$ and $u_{0}=(m_{\mathrm{o}}+\Delta_{\mathrm{o}})^{2}$, where the subscripts i and o represent the incoming and outcoming mesons, respectively. The values of $\Delta_{\mathrm{i}}$ and $\Delta_{\mathrm{o}}$ should be smaller than the experimental value of the distance between the ground and first excited state. In general, the $\Delta_{\mathrm{i}}$ and $\Delta_{\mathrm{o}}$ are taken as 0.5 GeV.

\begin{figure}[htbp]
    \centering
    \subfigure[]{
    \begin{minipage}[t]{4cm}
       \centering
       \includegraphics[width=4cm]{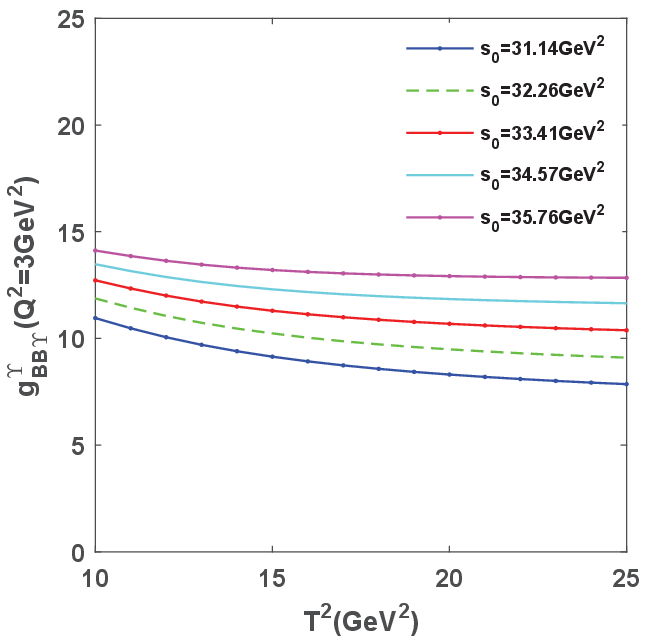}
    \end{minipage}
    }
   \subfigure[]{
   \begin{minipage}[t]{4cm}
       \centering
       \includegraphics[width=4cm]{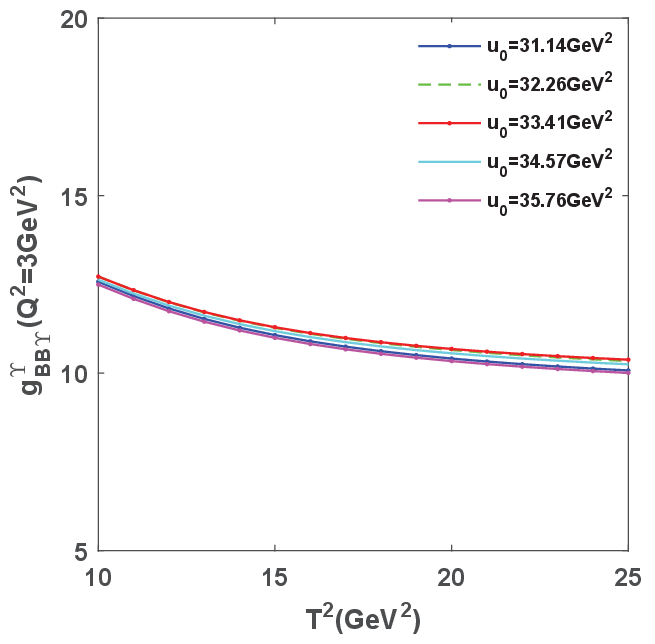}
    \end{minipage}
}
\caption{The strong coupling constants $g_{BB\Upsilon}^{\Upsilon}$ on Borel parameter $T^{2}$ in different values of $s_{0}$(a) and $u_{0}$(b).}
\label{fig:su}
\end{figure}

\begin{figure}[htbp]
\centering
\includegraphics[width=6cm]{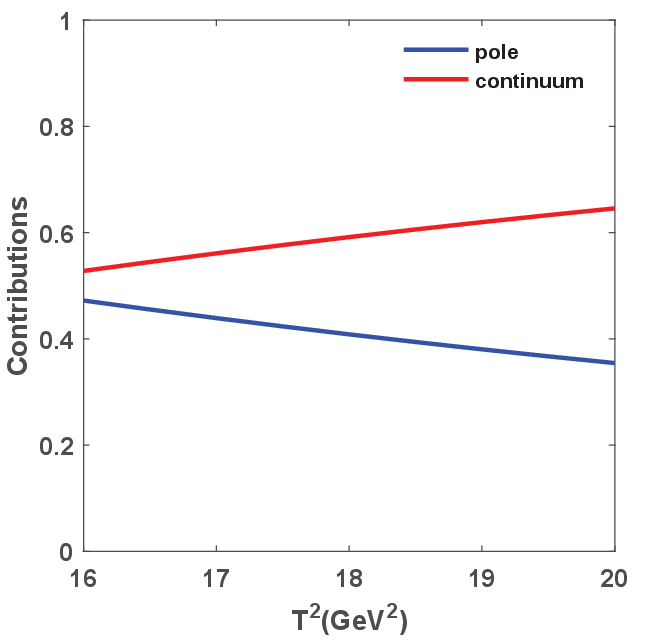}
\caption{The pole and continuum contributions with variation of the Borel parameter $T^{2}$ for the vertex $BB\Upsilon$ in $\Upsilon$ off-shell case.}
\label{fig:PC}
\end{figure}

\begin{figure*}[htbp]
\centering
\includegraphics[width=16cm]{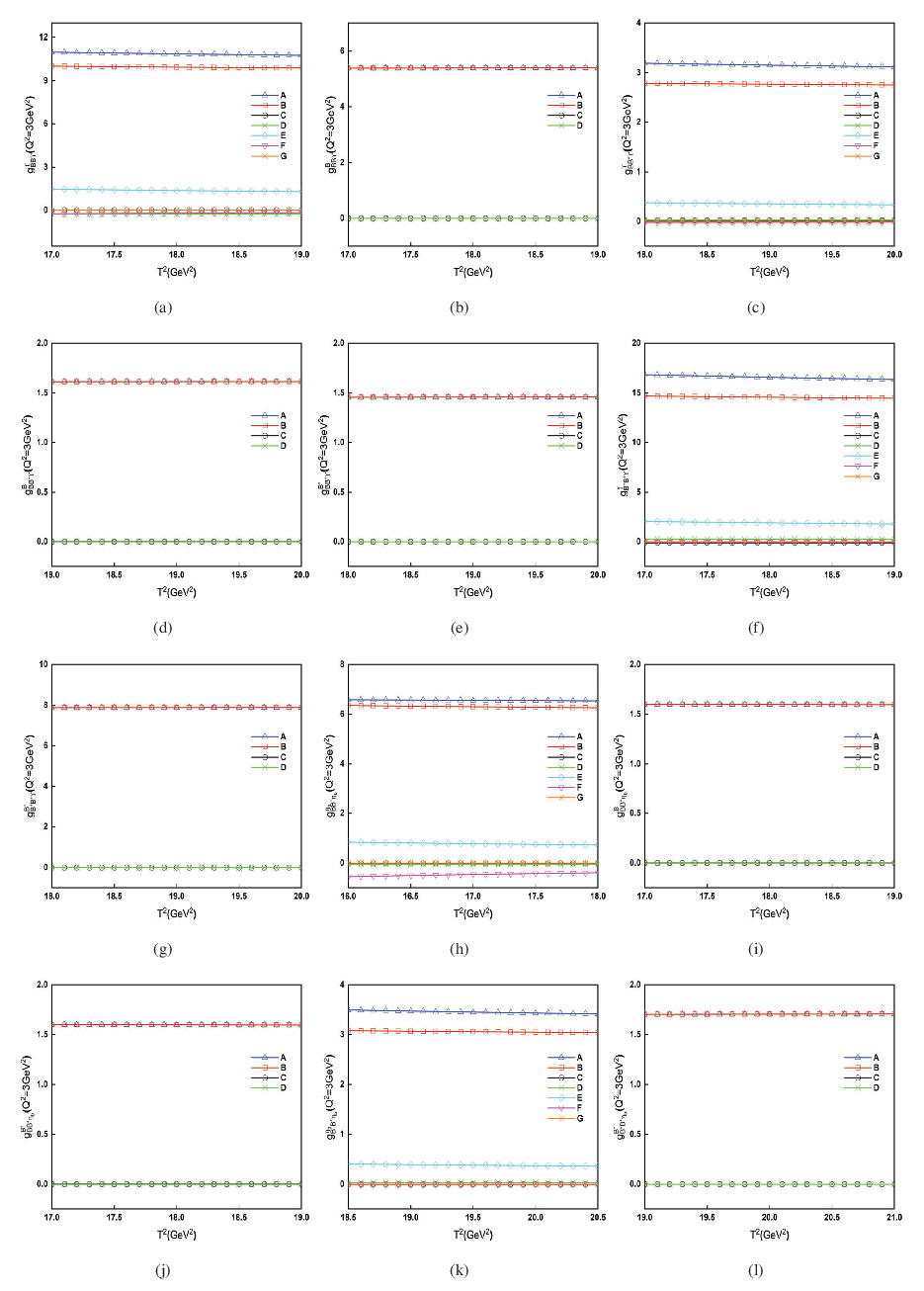}
\caption{The contributions of different vacuum condensate terms with variations of the Borel parameter $T^{2}$ for $BB\Upsilon$ (a-b), $BB^{*}\Upsilon$ (c-e), $B^{*}B^{*}\Upsilon$ (f-g), $BB^{*}\eta_{b}$ (h-j), $B^{*}B^{*}\eta_{b}$ (k-l), where A-G denote the total, perturbative term, $\langle g_{s}^{2}G^{2}\rangle, \langle f^{3}G^{3}\rangle,\langle\overline{q}q\rangle, \langle\overline{q}g_{s}\sigma G q\rangle$, and $\langle\overline{q}q\rangle\langle g_{s}^{2}G^{2}\rangle$ contributions.}
\label{BW}
\end{figure*}

Taking strong coupling constant $g_{BB\Upsilon}^{\Upsilon}$ as an example, we introduce how to choose these above parameters. Fixing $Q^{2}=3$ GeV$^{2}$ in Eq. (\ref{eq:20}), we firstly plot the strong coupling constant $g_{BB\Upsilon}^{\Upsilon}$ on Borel parameter $T^{2}$ in different values of $s_{0}$ and $u_{0}$ and this dependence is shown in Fig. \ref{fig:su}, which indicates that the results have good stability when $s_{0}$ and $u_{0}$ are taken as 31.14 $\sim$ 35.76 GeV$^{2}$ (the corresponding values of $\Delta_{\mathrm{i}}$ and $\Delta_{\mathrm{o}}$ are 0.3 $\sim$ 0.7 GeV). The pole and continuum contributions with variation of the Borel parameter $T^{2}$ are shown in Fig. \ref{fig:PC}. In can be seen from this figure that the condition of pole dominance($\geq40\%$) is satisfied in the range 17 GeV$^{2}$$\leq T^{2}\leq$ 19 GeV$^{2}$. The contributions of the total, perturbative and all vacuum condensate terms for vertex $BB\Upsilon$ are explicitly shown in Fig. \ref{BW}(a-b). These results showed good stability with variation of Borel parameter $T^{2}$ in the range 17 GeV$^{2}$$\leq T^{2}\leq$ 19 GeV$^{2}$, which indicates the condition of OPE convergence is also satisfied. This flat region(17 GeV$^{2}$$\leq T^{2}\leq$ 19 GeV$^{2}$) is usually called the 'Borel Window' that is used to extract the final results. According to similar analysis with $g_{BB\Upsilon}^{\Upsilon}$, the Borel Window for other strong vertices are also determined with $\Delta_{\mathrm{i}}=\Delta_{\mathrm{o}}=0.5$ GeV, which are also shown in Fig. \ref{BW}.

\begin{figure}[htbp]
\centering
\includegraphics[width=6cm]{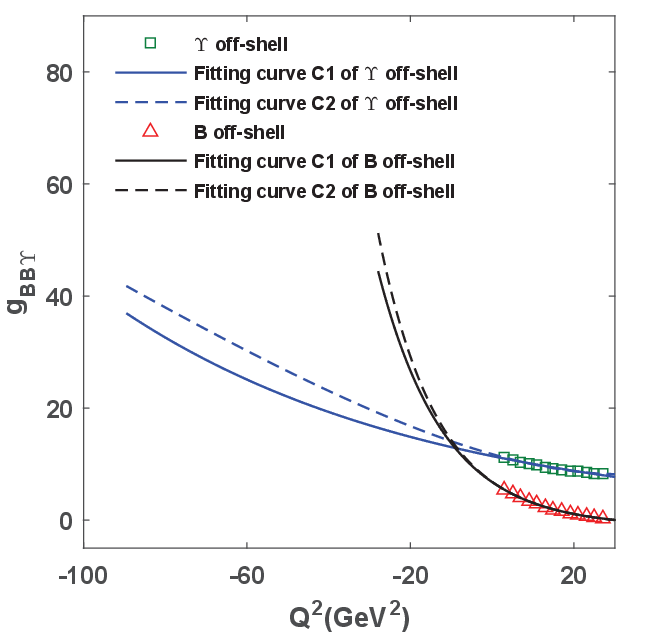}
\caption{The Fitting curves of the coupling constants for vertex $BB\Upsilon$ in $\Upsilon$ and $B$ off-shell. }
\label{fig:fc12}
\end{figure}

By taking different values of $Q^{2}$, the momentum dependent strong coupling constant $g(Q^{2})$ can be obtained with the values of $Q^{2}$ being taken as 3 $\sim$ 28 GeV$^{2}$ uniformly.
As for the on-shell values of these coupling constants, they will be obtained by extrapolating the results($g(Q^{2})$) into the time-like regions($Q^{2}<0$). This process is implemented by fitting $g(Q^{2})$ with appropriate functions and setting the intermediate meson on-shell($Q^{2}=-m^{2}_{\mathrm{on-shell}}$). In order to choose appropriate fitting functions, two conditions should be satisfied. The first one is that the strong coupling constants should be well fitted in the space-like regions($Q^{2}>0$), another is that the values of the strong coupling constants in deep time-like regions should converge. In addition, the on-shell values for the same vertex in different off-shell cases should be as close with each other as possible. Considering these above requirements, the uncertainties of the fitting function selection will be greatly reduced. Taking the strong coupling constant $g_{BB\Upsilon}$ as an example, $g(Q^{2})$ can be fitted into two functions with $\Upsilon$ and $B$ being off-shell cases, which are expressed as follows,
\begin{eqnarray}
\notag
C1:&&g_{BB\Upsilon }^\Upsilon ({Q^2}) = 11.43{e^{ - 0.013{Q^2}}}\\
\notag
&&g_{BB\Upsilon }^B({Q^2}) = 7.99{e^{ - 0.062{Q^2}}} - 1.20\\
\notag
C2:&&g_{BB\Upsilon }^\Upsilon ({Q^2}) = 11.78{e^{0.02{Q^2}}} - 0.45{Q^2}\\
&&g_{BB\Upsilon }^B({Q^2}) = 6.84{e^{ - 0.072{Q^2}}} - 0.03{Q^2}
\end{eqnarray}
Their fitting curves are shown in Fig. \ref{fig:fc12}, where we can see that both the fitting curves $C1$ and $C2$ are well fitted in the space-like regions for $\Upsilon$ and $B$ off-shell cases, and the on-shell values are well convergent. For the case of $g_{BB\Upsilon}^{\Upsilon}$, the curve $C1$(the blue solid line) is better fitted in the space-like regions than curve $C2$(the blue dashed line). Thus, fitting function $C1$ is selected for $g_{BB\Upsilon}$ in $\Upsilon$ off-shell case. For $B$ off-shell case, the curves $C1$ and $C2$ almost overlap with each other in the space-like regions (the black solid and dashed lines). As stated above, the on-shell values for same vertex in different off-shell cases should be close to each other. We can see that curve $C1$ is more appropriate for this condition. Thus, $C1$ is selected to fit $g_{BB\Upsilon}$ in $B$ off-shell case.

\begin{figure}[htbp]
\centering
\includegraphics[width=6cm]{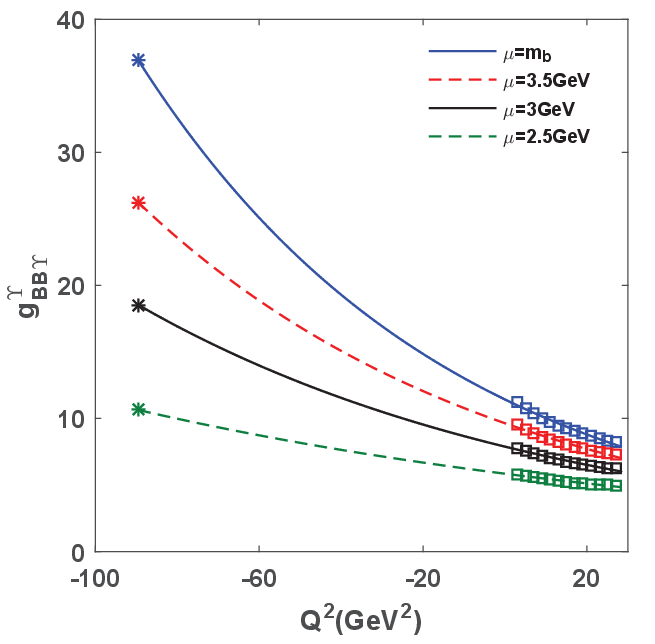}
\caption{The dependence of strong coupling constant $g_{BB\Upsilon}^{\Upsilon}$ on bottom quark mass with different energy-scale.}
\label{fig:scale}
\end{figure}

We know that the mass of bottom quark is strongly energy-scale dependent, which will also influence the final results obtained by QCDSR. For coupling constant $g_{BB\Upsilon}^{\Upsilon}(Q^{2})$ as an example, we plot this dependence in Fig. \ref{fig:scale} whose values are fitted by $C1$ function. From this figure, we can see that the results are indeed strongly dependent on energy-scale $\mu$. Actually, it is difficult for us to determine the appropriate energy-scale when calculating the coupling constants with QCDSR or LCSR. In some previous works, similar research works were carried out by the QCDSR and LCSR with different energy-scales of heavy quarks\cite{Wang:2013iia,Khodjamirian:2020mlb}. In the present work, the strong coupling constants will be obtained with the mass of bottom quark in energy-scale $\mu=m_{b}$.

\begin{figure}[H]
    \centering
    \subfigure[]{
    \begin{minipage}[t]{4cm}
       \centering
       \includegraphics[width=4.5cm]{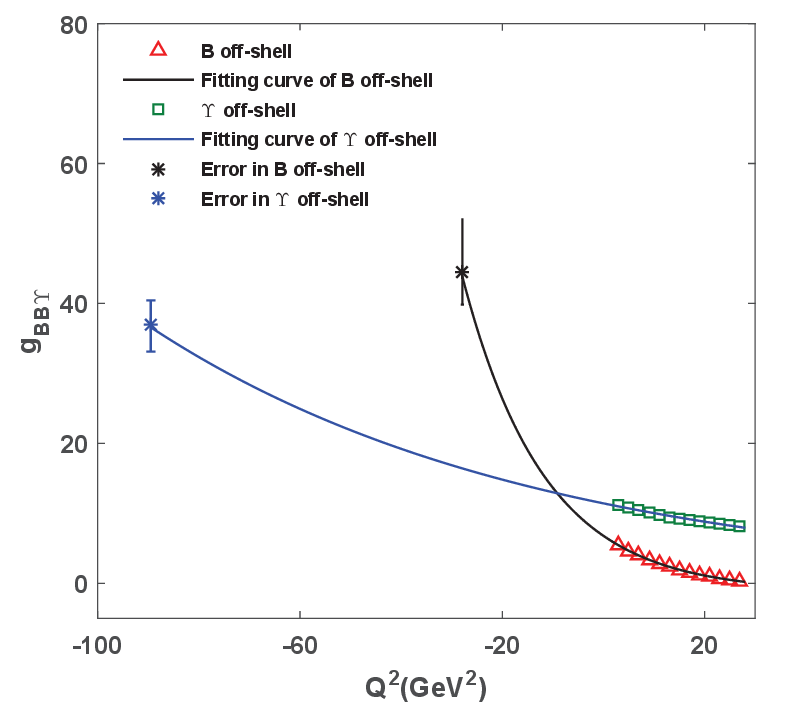}
    \end{minipage}
    }
   \subfigure[]{
   \begin{minipage}[t]{4cm}
       \centering
       \includegraphics[width=4.5cm]{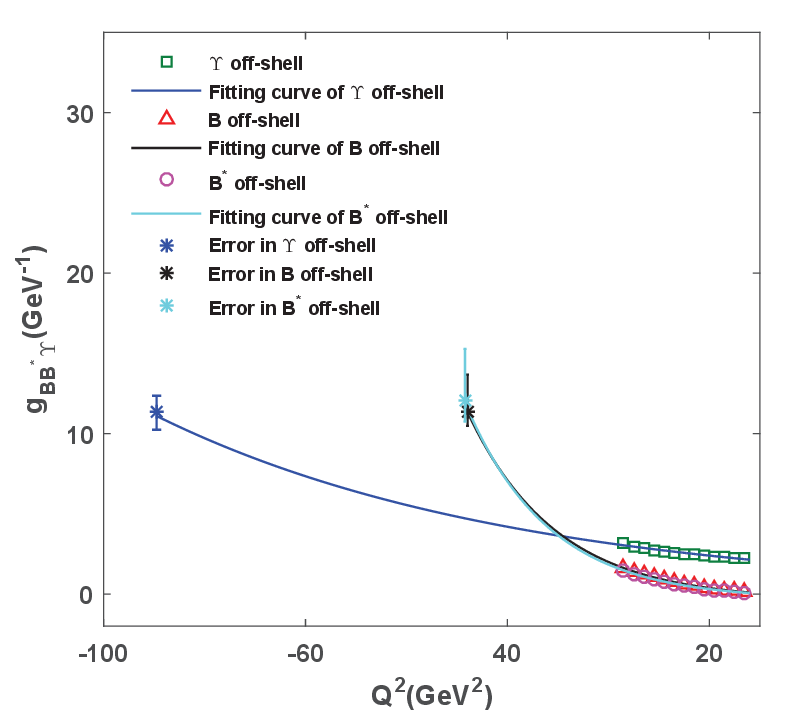}
    \end{minipage}
}
\subfigure[]{
   \begin{minipage}[t]{4cm}
       \centering
       \includegraphics[width=4.5cm]{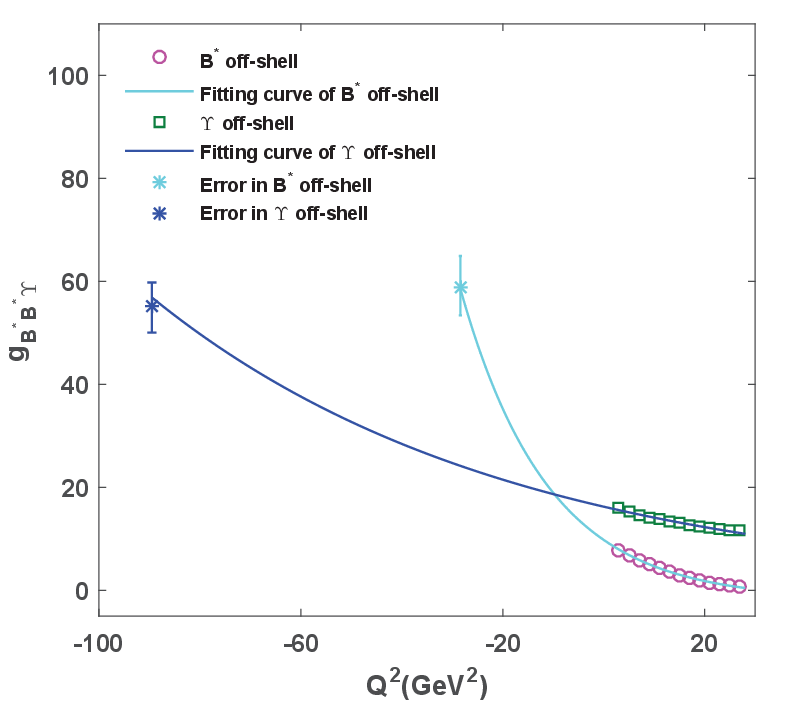}
    \end{minipage}
}
\subfigure[]{
    \begin{minipage}[t]{4cm}
       \centering
       \includegraphics[width=4.5cm]{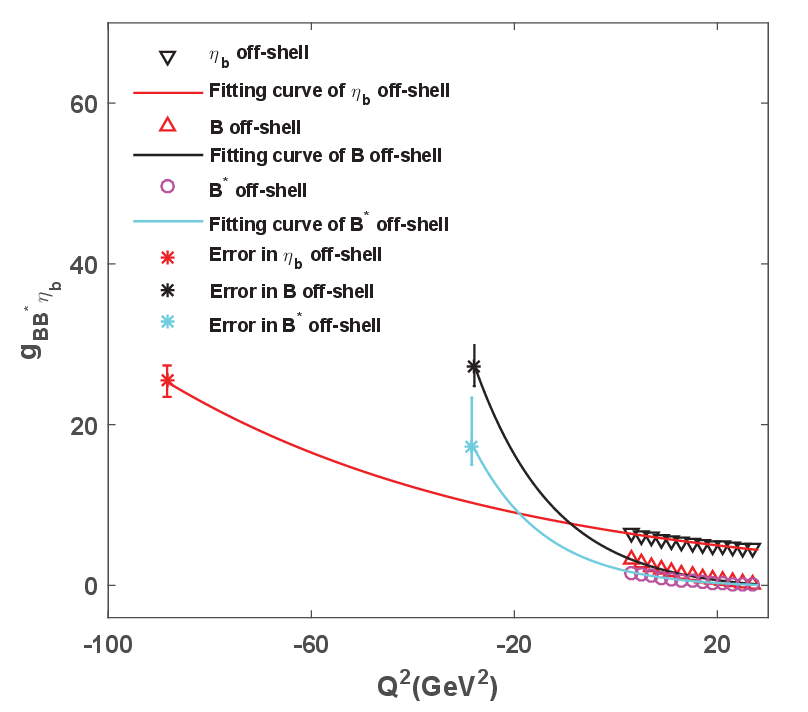}
    \end{minipage}
    }
   \subfigure[]{
   \begin{minipage}[t]{4cm}
       \centering
       \includegraphics[width=4.5cm]{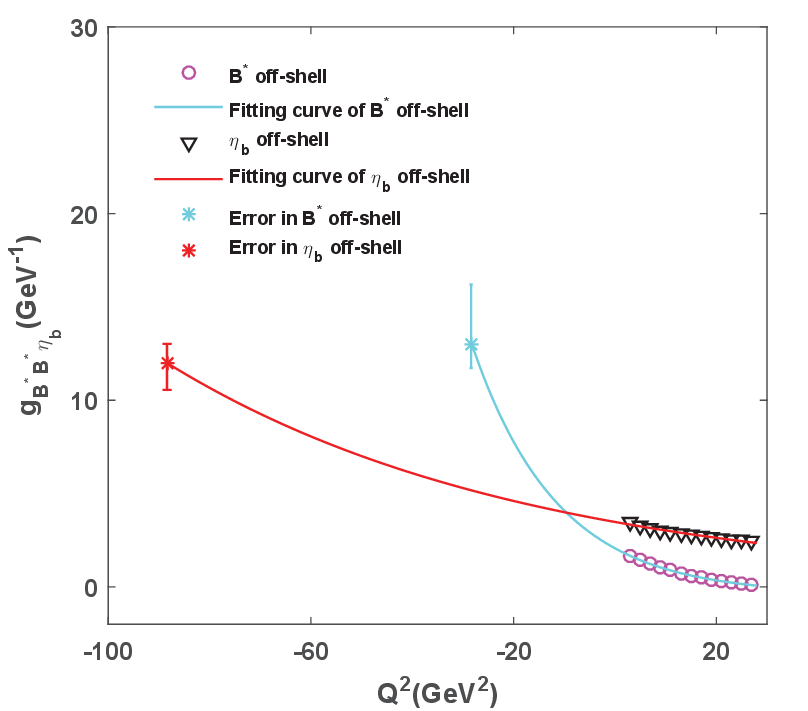}
    \end{minipage}
}
\caption{The fitting curves of coupling constants for vertices $BB\Upsilon$ (a), $BB^{*}\Upsilon$(b), $B^{*}B^{*}\Upsilon$(c), $BB^{*}\eta_{b}$(d) and $B^{*}B^{*}\eta_{b}$(e). In these figures, all possible off-shell cases are considered.}
\label{fig:FF}
\end{figure}

Finally, the momentum dependent strong coupling constants can be uniformly fitted into the following analytical function,
\begin{eqnarray}\label{eq:26}
g(Q^{2})=Fe^{-GQ^{2}}+H
\end{eqnarray}
where the values of parameters $F$, $G$ and $H$ are shown in Table~\ref{FFSC}. The fitting diagrams of strong coupling constants for each vertex are shown in Fig. \ref{fig:FF}.

Then, $g(Q^{2})$ are extrapolated into the time-like regions $(Q^{2}<0)$ by Eq. (\ref{eq:26}), and on-shell condition is satisfied by setting $Q^{2}=-m_{\mathrm{on-shell}}^{2}$. The on-shell values of strong coupling constants for different off-shell cases are obtained and are listed in the last column of Table~\ref{FFSC}.

\begin{table}[htbp]
\begin{ruledtabular}\caption{The fitting parameters in Eq. (\ref{eq:26}) and the strong coupling constants for different vertices and off-shell cases.}
\label{FFSC}
\begin{tabular}{l l l l l l}
Vertex&off-shell&$F$&$G$&$H$&$g(Q^{2}=-m_{\mathrm{on-shell}}^{2})$ \\ \hline
\multirow{2}*{$BB\Upsilon$}&$\Upsilon$&$11.43$&$0.013$&$0$&$36.92^{+3.51}_{-3.80}$  \\
~&$B$ &$7.99$&$0.062$&$-1.20$&$44.42^{+11.59}_{-4.60}$   \\  \hline
\multirow{3}*{$BB^{*}\Upsilon$}&$\Upsilon$ &$3.17$&$0.014$&$0$&$11.35^{+1.01}_{-1.11}$GeV$^{-1}$  \\
~&$B$ &$2.39$&$0.057$&$-0.39$&$11.36^{+2.31}_{-0.88}$GeV$^{-1}$   \\
~&$B^{*}$ &$2.17$&$0.061$&$-0.33$&$12.04^{+3.24}_{-1.29}$GeV$^{-1}$   \\ \hline
\multirow{2}*{$B^{*}B^{*}\Upsilon$}&$\Upsilon$ &$16.24$&$0.014$&$0$&$55.10^{+4.67}_{-5.07}$  \\
~&$B^{*}$ &$11.63$&$0.058$&$-1.84$&$58.93^{+5.98}_{-5.55}$   \\  \hline
\multirow{3}*{$BB^{*}\eta_{b}$}&$\eta_{b}$ &$6.72$&$0.015$&$0$&$25.52^{+1.84}_{-2.06}$  \\
~&$B$ &$4.71$&$0.064$&$-0.66$&$27.31^{+6.35}_{-2.51}$   \\
~&$B^{*}$ &$2.39$&$0.071$&$-0.28$&$17.34^{+6.03}_{-2.34}$   \\ \hline
\multirow{2}*{$B^{*}B^{*}\eta_{b}$}&$\eta_{b}$ &$3.48$&$0.014$&$0$&$11.97^{+1.05}_{-1.41}$GeV$^{-1}$  \\
~&$B^{*}$ &$2.46$&$0.060$&$-0.39$&$13.00^{+3.20}_{-1.27}$GeV$^{-1}$   \\
\end{tabular}
\end{ruledtabular}
\end{table}

For each vertex, the on-shell values of coupling constants for different off-shell cases should be equal to each other. For vertex $BB^{*}\Upsilon$ as an example, their central values for different off-shell cases are $g_{BB^{*}\Upsilon}^{\Upsilon}=11.35$ GeV$^{-1}$, $g_{BB^{*}\Upsilon}^{B}=11.36$ GeV$^{-1}$ and $g_{BB^{*}\Upsilon}^{B^{*}}=12.04$ GeV$^{-1}$, which are consistent well with each other. Thus, it is reasonable to determine the final values of the strong coupling constants by taking their average values. Finally, the results of the strong coupling constants for different strong vertices are determined as,
\begin{eqnarray}
\notag
&&g_{BB\Upsilon}=40.67^{+7.55}_{-4.20} \\
\notag
&&g_{BB^{*}\Upsilon}=11.58^{+2.19}_{-1.09} \mathrm{GeV}^{-1} \\
\notag
&&g_{B^{*}B^{*}\Upsilon}=57.02^{+5.32}_{-5.31} \\
\notag
&&g_{BB^{*}\eta_{b}}=23.39^{+4.74}_{-2.30} \\
&&g_{B^{*}B^{*}\eta_{b}}=12.49^{+2.12}_{-1.35} \mathrm{GeV}^{-1}
\end{eqnarray}

\begin{large}
\section{Conclusions}\label{sec4}
\end{large}

In this work, we systematically analyze the strong vertices $BB\Upsilon$, $BB^{*}\Upsilon$, $B^{*}B^{*}\Upsilon$, $BB^{*}\eta_{b}$, $B^{*}B^{*}\eta_{b}$ using the QCD sum rules, where all off-shell cases are considered for each vertex. Under this physical sketch, the momentum dependent coupling constants are firstly obtained in the space-like ($Q^{2}>0$) regions, then they are fitted into appropriate analytical functions. By extrapolating these functions into the time-like ($Q^{2}<0$) regions and taking $Q^{2}=-m^{2}_{\mathrm{on-shell}}$, we obtain the on-shell strong coupling constants. For each vertex, we take the average value of the on-shell strong coupling constants for all off-shell cases as the final results. These coupling constants are valuable in describing the dynamical behaviors of hadrons. For example, these coupling constants are important input parameters to analyze the final-state interactions in the heavy quarkonium decays, or to calculate the absorption cross sections in understanding the heavy quarkonium absorptions in hadronic matter.

\section*{Acknowledgements}

This paper has been presented on the web site \href{https://arxiv.org/pdf/2307.05090.pdf}{https://arxiv.org/pdf/2307.05090.pdf}. This project is supported by National Natural Science Foundation, Grant Number 12175068 and Natural Science Foundation of HeBei Province, Grant Number A2018502124.

\end{document}